# Pressure-Induced Superconductivity in Topological Heterostructure (PbSe)$_5$(Bi$_2$Se$_3$)$_6$

Cuiying Pei[1#], Peng Zhu[2,3,4#], Bingtan Li[5,6], Yi Zhao[1], Lingling Gao[1], Changhua Li[1], Shihao Zhu[1], Qinghua Zhang[7], Tianping Ying[7], Lin Gu[7], Bo Gao[8], Huiyang Gou[8], Yansun Yao[9], Jian Sun[10], Hanyu Liu[5,6], Yulin Chen[1,11,12], Zhiwei Wang[2,3,4*], Yugui Yao[2,3], Yanpeng Qi[1,11,13*]

1. School of Physical Science and Technology, ShanghaiTech University, Shanghai 201210, China
2. Centre for Quantum Physics, Key Laboratory of Advanced Optoelectronic Quantum Architecture and Measurement (MOE), School of Physics, Beijing Institute of Technology, Beijing 100081, China
3. Beijing Key Lab of Nanophotonics and Ultrafine Optoelectronic Systems, Beijing Institute of Technology, Beijing 100081, China
4. Material Science Center, Yangtze Delta Region Academy of Beijing Institute of Technology, Jiaxing, 314011, P. R. China
5. State Key Laboratory of Superhard Materials and International Center for Computational Method and Software, College of Physics, Jilin University, Changchun 130012, China
6. International Center of Future Science, Jilin University, Changchun 130012, China
7. Beijing National Laboratory for Condensed Matter Physics, Institute of Physics, Chinese Academy of Sciences, Beijing 100190, China
8. Center for High Pressure Science and Technology Advanced Research, Beijing, 100094, China
9. Department of Physics and Engineering Physics, University of Saskatchewan, Saskatoon, Saskatchewan S7N 5E2, Canada
10. National Laboratory of Solid State Microstructures, School of Physics and Collaborative Innovation Center of Advanced Microstructures, Nanjing University, Nanjing 210093, China
11. ShanghaiTech Laboratory for Topological Physics, ShanghaiTech University, Shanghai 201210, China
12. Department of Physics, Clarendon Laboratory, University of Oxford, Parks Road, Oxford OX1 3PU, UK
13. Shanghai Key Laboratory of High-resolution Electron Microscopy, ShanghaiTech University, Shanghai 201210, China

# These authors contributed to this work equally.
* Correspondence should be addressed to Y.Q. (qiyp@shanghaitech.edu.cn) or Z.W. (zhiweiwang@bit.edu.cn)

Recently, the natural heterostructure of $(PbSe)_5(Bi_2Se_3)_6$ has been theoretically predicted and experimentally confirmed as a topological insulator. In this work, we induce superconductivity in $(PbSe)_5(Bi_2Se_3)_6$ by implementing high pressure. As increasing pressure up to 10 GPa, superconductivity with $T_c \sim 4.6$ K suddenly appears, followed by an abrupt decrease. Remarkably, upon further compression above 30 GPa, a new superconducting state arises, where pressure raises the $T_c$ to an unsaturated 6.0 K within the limit of our research. Combining XRD and Raman spectroscopies, we suggest that the emergence of two distinct superconducting states occurs concurrently with the pressure-induced structural transition in this topological heterostructure $(PbSe)_5(Bi_2Se_3)_6$.

Heterostructures are layered structures that contain an interface between different materials, which is often advantageous to engineer the electronic energy bands in many solid-state device applications, including semiconductor lasers, solar cells, and transistors.[1-8] Fabrication of heterostructures has long been limited by deposition or epitaxial techniques, eg, molecular beam epitaxy (MBE) and metalorganic chemical vapor deposition (MOCVD), hindering extensive studies of the unique material systems. Hence, natural multilayer heterostructures with homogeneous interfaces are desired and can provide new opportunities to study novel physical properties.

The Pb-based homologous series of $(PbSe)_5(Bi_2Se_3)_{3m}$ (m = 1, 2), one of the natural multilayer heterostructures, has recently attracted attention.[9-13] The structure is generally regarded as an alternating layered structure of $m$ quintuple layers (QLs) of $Bi_2Se_3$ with bilayer PbSe, and it is naturally formed as a multilayer heterostructure. Because the binary compound $Bi_2Se_3$ is well known to be a topological insulator (TI)[14,15] and PbSe is a topologically trivial one[16], $(PbSe)_5(Bi_2Se_3)_{3m}$ (m = 1, 2) consists of ultrathin TI layers separated by trivial-insulator layers. Interestingly, Nakayama et al. have observed gapped Dirac-cone dispersions with a large band gap of 0.5 eV for m = 2 *via* angle-resolved photoemission spectroscopy (ARPES), showing that topological interface states are effectively encapsulated by the PbSe block layers.[11] Such a multilayer system with topological and ordinary insulating layers is an interesting playground for realizing novel topology.

Recently, unconventional superconductivity has been reported experimentally in $(PbSe)_5(Bi_2Se_3)_{3m}$ (m = 1, 2) by Cu intercalations and Ag doping, offering a potential platform to observe the Majorana fermion state at the boundary of natural heterostructures.[13, 17, 18] Pressure is an effective method to tune the lattice structure and to manipulate electronic state without introducing impurities. It has been shown that pressure can induce superconductivity in topological phases of matter.[19-21] In this work, we report the pressure-induced two distinct superconducting states in topological Heterostructure $(PbSe)_5(Bi_2Se_3)_6$. Under high pressures, superconductivity with $T_c \sim$ 4.6 K suddenly appears at around 10 GPa and is then suppressed abruptly, forming an SC-I in the pressure-temperature phase diagram. Another SC-II emerges above 30 GPa, of which the $T_c$ increases slowly upon compression and a maximum and unsaturated $T_c$ of 6.0 K is obtained within the limit of our research. The pressure-induced two distinct SC in topological heterostructure $(PbSe)_5(Bi_2Se_3)_6$ is accompanied by a structural transition, as evidenced by both the XRD and Raman data.

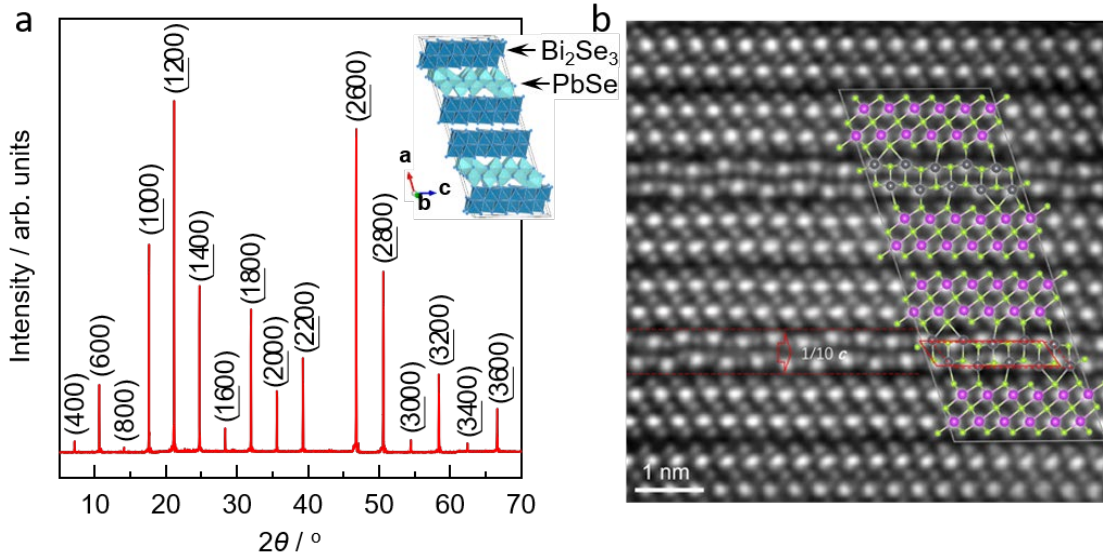

Figure 1. (a) XRD pattern of $(PbSe)_5(Bi_2Se_3)_6$ single crystal at ambient pressure. Insert: crystal structure of $(PbSe)_5(Bi_2Se_3)_6$. (b) Angular dark-field (ADF) image of $(PbSe)_5(Bi_2Se_3)_6$. Their respective crystal structures are superimposed. Aberration-corrected scanning transmission electron microscopy (STEM) images of $(PbSe)_5(Bi_2Se_3)_6$ with well-organized atomic position along [010]. The crystal structures given in the Inorganic Crystal Structure Database (ICSD) are superimposed for comparison. Solid and broken parallelograms highlight the supposed and actual atomic position of the PbSe layer, respectively. The red broken lines isolate the PbSe layer that drifts along the *c*-axis by 1/10.

$(PbSe)_5(Bi_2Se_3)_6$ crystallizes in a monoclinic structure $C2/m$ (No. 12). Prior to high-pressure measurements, we first checked the sample quality by single-crystal XRD diffractions and scanning transmission electron microscopy (STEM), as shown in Figures 1. The single-crystal XRD data reveals that the (*h*00) plane is a natural cleavage facet of as-grown single crystals. The extracted lattice parameters are $c$ = 50.3858 Å, in agreement with previous reports.[22] Annular dark-field scanning transmission electron microscopy (ADF-STEM) images of $(PbSe)_5(Bi_2Se_3)_6$ along [010] are shown in Figure 1b. Bi atoms are located at the centre of distorted Se octahedra, while Pb atoms are surrounded by Se atoms and form a distorted polyhedron. An alternate stack of $Bi_2Se_3$ and PbSe along the *a* axis builds a multilayer heterostructure in a monoclinic unit cell belonging to the space group of *C*2/*m*. An interesting observation is that one slice of the PbSe layer is drifted 1/10 off the *c*-axis (highlighted in the red solid and broken parallelograms). These atomic drifts within one unit cell cannot be reflected in our powder x-ray diffraction patterns, but they are prevalent in all the observed ADF-STEM images. Such dislocations should be an inherent property of this misfit layered compound, introducing significant strain and distortion to alter its physical properties. The above characterizations indicate the high quality of our samples.

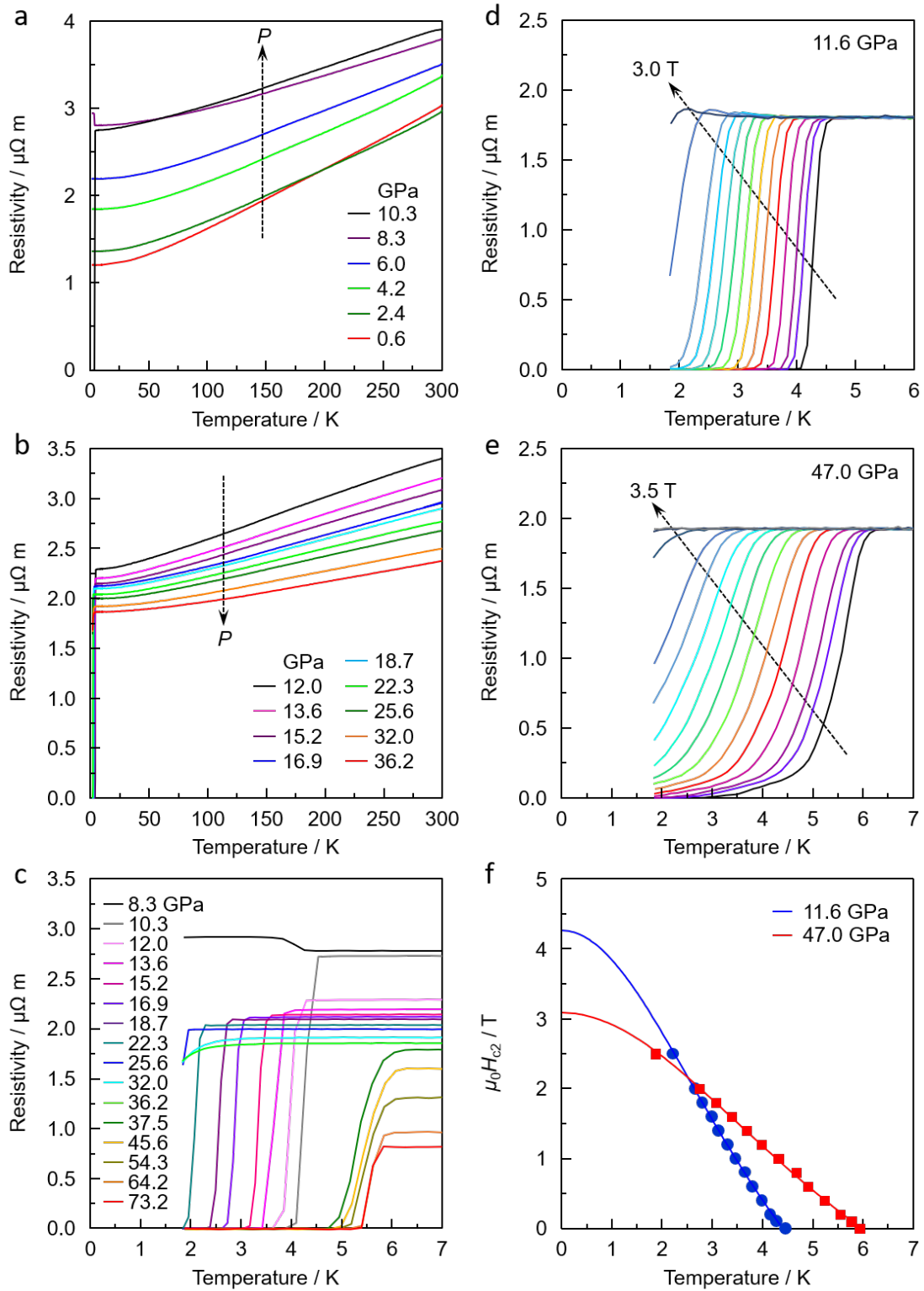

Figure 2. Transport properties of $(PbSe)_5(Bi_2Se_3)_6$ as a function of pressure. Electrical resistivity as a function of temperature for pressures of 0.6 - 10.3 GPa (a) and 12.0 - 36.2 GPa (b). (c) Temperature-dependent resistivity of $(PbSe)_5(Bi_2Se_3)_6$ in the vicinity of the superconducting transition. Temperature dependence of resistivity under different magnetic fields for $(PbSe)_5(Bi_2Se_3)_6$ at 11.6 GPa (d) and 47.0 GPa (e). (f) Temperature dependence of upper critical

field $\mu_0H_{c2}(T)$ at 11.6 GPa and 47.0 GPa. Here, the $T_c$s are determined at 90 % of the normal state resistivity just above the onset superconducting transition temperature. The solid dots and square stand for the temperature dependence of resistivity under different magnetic fields for $(PbSe)_5(Bi_2Se_3)_6$ at 11.6 GPa and 47.0 GPa, respectively. The solid lines represent the fits using the Ginzburg-Landau formula.

Figure 2 shows the evolution of temperature dependence of the electrical resistivity of $(PbSe)_5(Bi_2Se_3)_6$ for pressure up to 73.4 GPa. At 0.6 GPa, the resistivity of $(PbSe)_5(Bi_2Se_3)_6$ in the whole pressure range shows a metallic-like nature with residual resistivity ratio RRR = 2.53. In the low-pressure region, increasing pressure initially induces a continuous enhancement of the overall magnitude of $\rho$, with a maximum occurring at around 10 GPa. Subsequently, resistivity decreases slowly. Meanwhile, a superconducting transition where resistivity reaches zero emerges at 4.6 K suddenly appears. Subsequently, the superconducting transition temperature $T_c$ is suppressed to a minimum of 1.9 K at around 25-30 GPa. Surprisingly, $T_c$ starts to increase rapidly with further increases in pressure above 32 GPa, reaching a value of 6.0 K at 37.5 GPa. So, a pressure-induced reentrant superconducting state was observed for $(PbSe)_5(Bi_2Se_3)_6$. With the pressure further increasing, $T_c$ increases slowly to an unsaturated $T_c$ of 6.2 K obtained within the limit of our research.

There are two distinct pressure-induced superconducting regions. To further identify the difference between these two superconducting states, we applied magnetic fields on $(PbSe)_5(Bi_2Se_3)_6$ subjected to 11.6 and 47.0 GPa, respectively. As shown in Figure 2d, the zero-resistance state at 11.6 GPa is continuously suppressed by the magnetic field until it vanishes over 3.0 T. On application of the magnetic field to the compressed sample at 47.0 GPa, similar behavior was observed (Figure 2e). These results confirm the resistivity drop in both two SC states is related to superconducting transition. Figure 2f shows the G-L fitting on the field dependence of $T_c$ for $(PbSe)_5(Bi_2Se_3)_6$ at 11.6 and 47.0 GPa, respectively[23-25]. The estimated critical fields $\mu_0H_{c2}(0)$ ~ 4.3 and 3.1 T for 11.6 and 47.0 GPa, respectively. Although the $\mu_0H_{c2}$ obtained here is lower than its corresponding Pauli paramagnetic limit $H_P = 1.84T_c$, the slopes of $dH_{c2}/dT$ are notably different: −1.16 and −0.63 T/K for 11.6 and 47.0 GPa, respectively. Our results demonstrate distinct different nature between the two pressure-induced superconducting states. Using the relations $\xi_{GL}(0) = \sqrt{\phi_0/2\pi\mu_0 H_{c2}(0)}$, where $\phi_0$ is the flux quantum, we derive G-L coherence length $\xi_{GL}(0) = $ 9.8 nm at 11.6 GPa and 10.3 nm at 47.0 GPa, respectively. The details are summarized in Table I.

Table I. A summary of superconductivity properties of (PbSe)$_5$(Bi$_2$Se$_3$)$_6$ under high pressure.

| Phase | $T_c$ / K | $H_{c2}(0)$ / T | d$H_{c2}$/d$T$ / T/K | $\xi_{GL}(0)$ / nm |
| --- | --- | --- | --- | --- |
| SC-I (11.6 GPa) | 4.4 | 4.3 | -1.16 | 9.8 |
| SC-II (47.0 GPa) | 6.0 | 3.1 | -0.63 | 10.3 |

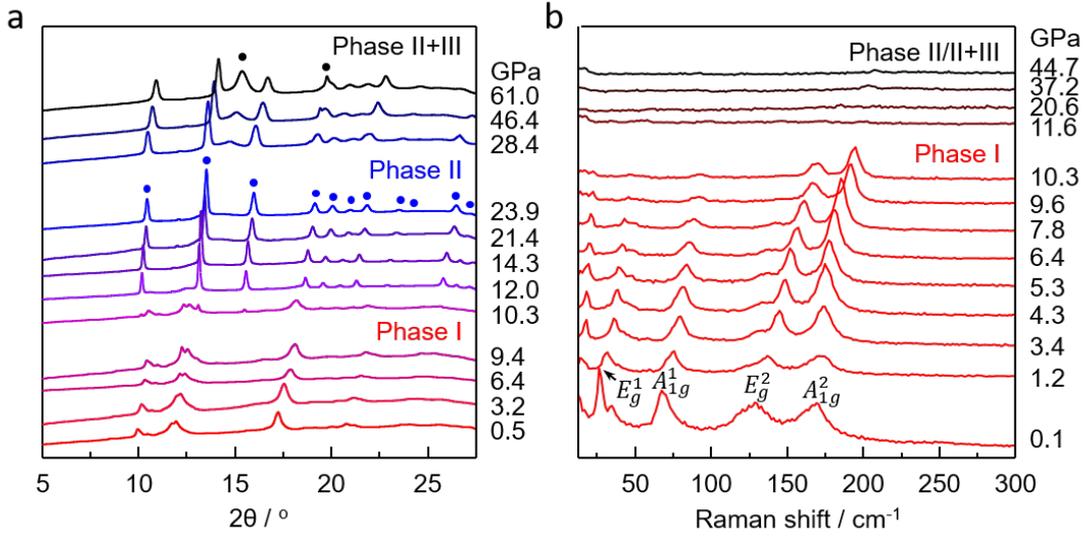

Figure 3. Pressure dependence of XRD patterns (a) and Raman spectra (b) at room temperature of (PbSe)$_5$(Bi$_2$Se$_3$)$_6$, respectively. High-pressure *in-situ* synchrotron XRD was measured in Shanghai Synchrotron Radiation Facility (SSRF) with x-ray wave-length λ = 0.6199 Å.

To investigate whether the observed two distinct superconducting states in pressurized (PbSe)$_5$(Bi$_2$Se$_3$)$_6$ are associated with a pressure-induced crystal structure phase transition, we performed *in situ* high-pressure XRD measurements. The XRD patterns of (PbSe)$_5$(Bi$_2$Se$_3$)$_6$ collected at different pressures are shown in Figure 3a. A representative refinement at 0.5 GPa is displayed in Figure S1 (supplementary material). All the diffraction peaks can be indexed well to the ambient monoclinic structure with a space group of *C*2/*m* (no. 12). Using ideal hydrostaticity Ne as pressure transmitting medium (PTM), the ambient phase is robust and could remain till 9.4 GPa. Above that, several peaks are involved which are attributed to a new high-pressure phase (Phase II). This phase is stable in a pressure range of 10.3-23.9 GPa. Above 28.4 GPa, Phase III was observed and coexisted with Phase II upon compression (Figures S2-S4). We could not get pure Phase III even at the pressure of 61.0 GPa. It should be noted that structure searches for this system by CALYPSO or other methods failed since its complicated atom configuration.[26-28] To derive more structural phase transition information, high-pressure *in-situ* Raman spectroscopy measurement was carried out on $(PbSe)_5(Bi_2Se_3)_6$. As shown in Figure 3b, four Raman active modes are clearly assigned, in which $E_g$ modes correspond to in-plane bond vibration and $A_{1g}$ modes reflect the out-of-plane one. With pressure increasing, all four phonon modes show a blue shift arising from an enhancement of the van der Walls forces between adjacent layers. An abrupt disappearance of Raman modes at 11.6 GPa indicates pressure-induced structural phase transition, which is consistent with our synchrotron XRD patterns. In summary, our *in-situ* XRD and Raman spectroscopy measurements demonstrate that two distinct high-pressure phases were achieved in $(PbSe)_5(Bi_2Se_3)_6$ upon compression.

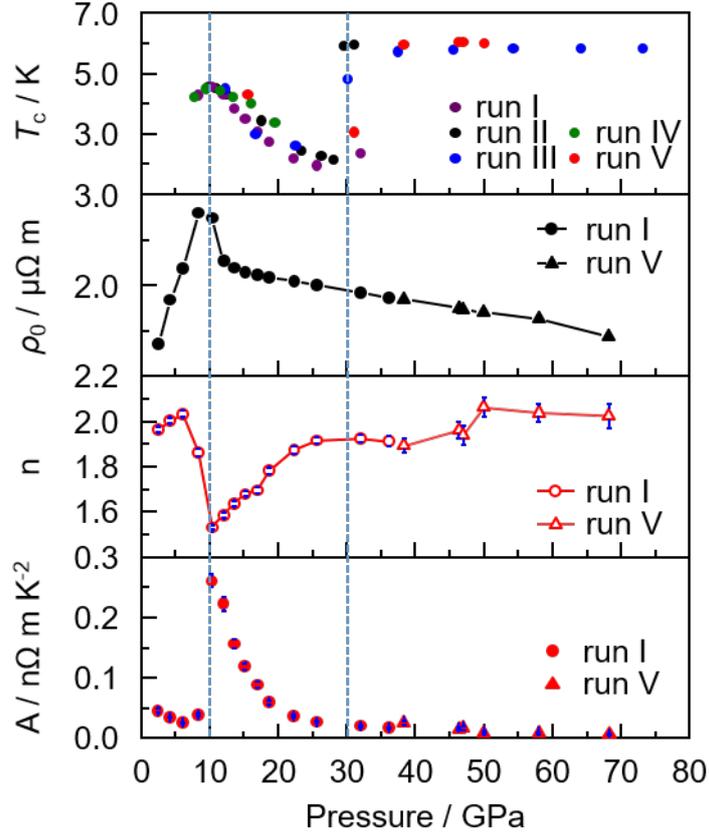

Figure 4. The temperature-pressure diagram of $(PbSe)_5(Bi_2Se_3)_6$. (a) Pressure-dependent $T_c$ of $(PbSe)_5(Bi_2Se_3)_6$. The red, olive, blue, purple, and black solid circles represent the $T_c$ extracted from different runs of resistivity measurements. Pressure-dependent residual resistivity $\rho_0$ (b), temperature exponent $n$ (c), and constant $A$ (d) of $(PbSe)_5(Bi_2Se_3)_6$.

The pressure-dependent $T_c$ and related physical parameters are summarized in Figure 4. To confirm the emergence of two distinct superconducting states under high pressure, we repeated the measurements with new samples for several runs and proved that all the results are reproducible (Figures S7-S11). The $P$–$T_c$ phase diagram reveals three different regions: the initial non-superconducting Phase I (around 0 ~ 10 GPa) and the pressure-induced superconducting Phase II (10 ~ 30 GPa) and Phase III (30 ~ 80 GPa). For the normal state, we fit the $\rho(T)$ curve by the equation $\rho = \rho_0 + AT^n$ from $T_c$ to 70 K, here $\rho_0$ is residual resistivity, and $A$ is constant. The pressure-dependent key transport parameters are also plotted in Figure 4. In Phase I, $\rho_0$ increases with pressure while $n$ maintains around 2, showing a Fermi-liquid behavior[29]. No superconductivity was observed down to 1.8 K in this pressure range. In Phase II, a superconducting transition with $T_c$ ~ 4.6 K suddenly appears accompanied by a first structural phase transition, however, $T_c$ is sensitive to the pressure and decreases sharply upon compression. Only a small drop without zero resistivity was observed at a pressure of around 30 GPa. Meanwhile, the $A$ quickly decreases from 0.25 nΩ m K$^{-2}$ to 0.01 nΩ m K$^{-2}$, also the $n$ increases from 1.5 to 2.0, implying that the electron-correlated states possibly transit from non-Fermi-liquid to Fermi-liquid state. In Phase III, a new superconducting phase suddenly emerges after the second structural phase transition, showing pressure-induced reemergence of superconductivity. $T_c$ was enhanced as high as 6 K, which is higher than that in phase II. Upon further increasing the pressure, $T_c$ increases slowly without saturation in our measured pressure limit. At this stage, the $A$ does not change so much and the $n$ keeps at around 2, which is similar to the initial phase.

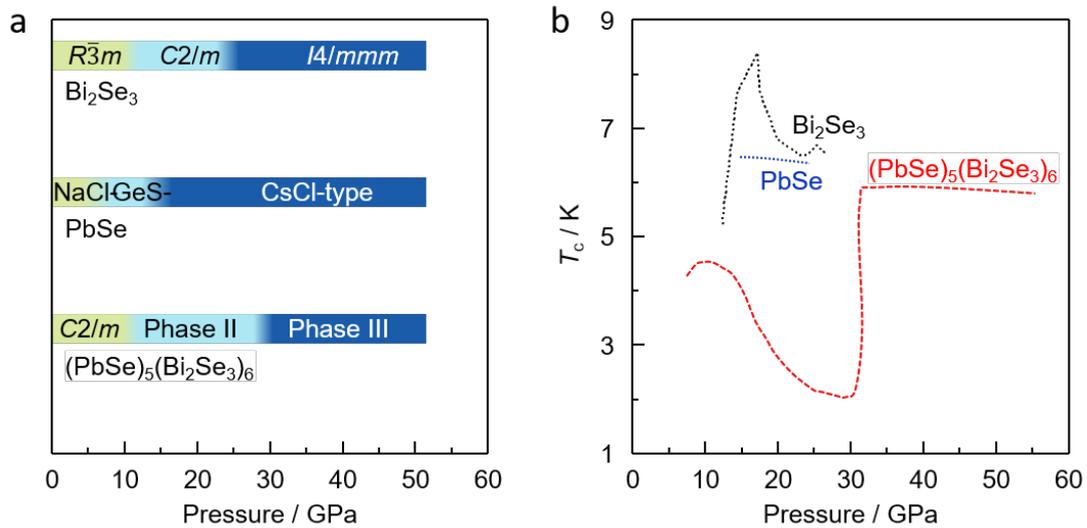

Figure 5. Evolution of crystal structure (a) and superconductivity (b) of heterostructure (PbSe)$_5$(Bi$_2$Se$_3$)$_6$ and building blocks (PbSe and Bi$_2$Se$_3$) under external pressure. [30-33, 15, 14, 34, 35]

Finally, we would like to compare the high-pressure effect between the multilayer heterostructure and building blocks (PbSe and Bi$_2$Se$_3$). First, pressure-induced multi-phase transitions were observed in heterostructures (PbSe)$_5$(Bi$_2$Se$_3$)$_6$ and its individual building blocks (PbSe and Bi$_2$Se$_3$). For PbSe, it undergoes a structural phase transition from NaCl- (Phase I) to GeS-type lattice (Phase II) at 2-6 GPa with resistivity abruptly rising.[30-32] In further compression above 15 GPa, another structural phase transition to a CsCl-type structure (Phase III) occurs, accompanied by a semiconductor-metal conversion. For Bi$_2$Se$_3$, a pressure-induced structural phase transition occurs from ambient rhombohedra structure (Phase I, $R\bar{3}m$) to monoclinic structure (Phase II, $C2/m$) at ~10 GPa and to body-centered tetragonal structure (Phase III, $I4/mmm$) at ~25 GPa.[15, 14] Second, the application of pressure effectively tuned the electronic properties, and superconductivity was induced in both heterostructures $(PbSe)_5(Bi_2Se_3)_6$ and building blocks (PbSe and $Bi_2Se_3$). For PbSe, superconductivity was only observed in phase III (~ 30 GPa) with $T_c$ of 6.5 K.[33]. Correspondingly $Bi_2Se_3$ becomes superconducting at around 12 GPa with $T_c$ of 4.4 K.[34] With pressure increasing, $T_c$ increases to a maximum of 8.2 K at 17 GPa and then decreases to 6.5 K at 23 GPa.[35] Alternating layers of rocksalt PbSe and $Bi_2Se_3$ present diverse symmetry and periodicity, which results in an incommensurate structure. Figure 5 summarizes the coevolution of crystal structures and $T_c$ under external pressure in $(PbSe)_5(Bi_2Se_3)_6$, PbSe, and $Bi_2Se_3$ for comparison. It is reasonable to predict the first superconducting dome given the onset of superconductivity in both PbSe and $Bi_2Se_3$ at moderate pressure. However, the reentrant of the second superconducting dome at phase III is quite unexpected. Such an emergent phenomenon might result from the interfacial effect of the infinite PbSe-$Bi_2Se_3$ stacking sequence. Further researches are highly desired with the primary goal of determining its high-pressure structure in the phase III.

In conclusion, we systematically investigate the electrical transport, structural, and lattice dynamical properties of topological heterostructure $(PbSe)_5(Bi_2Se_3)_6$ under high pressures up to 80 GPa. Application of pressure effectively tuned the electronic properties and crystal structure of $(PbSe)_5(Bi_2Se_3)_6$. Two pressure-induced superconducting states are observed upon compression, which is related to a structural transition as evidenced by both the XRD and Raman measurements. Besides natural multilayer heterostructure and nontrivial topology, our results demonstrate that $(PbSe)_5(Bi_2Se_3)_6$ exhibits new ground states upon compression.

**EXPERIMENTAL SECTION**

High-quality single crystals of $(PbSe)_5(Bi_2Se_3)_6$ were grown by using a combination of a melting method and modified Bridgman method[36]. High-purity starting materials of Bi, Pb, and Se are loaded in a quartz tube with the ratio of PbSe : $Bi_2Se_3$ = 38 : 62. The tube was sealed after it was evacuated to a vacuum of 2×10$^{-4}$ Pa. The raw materials are reacted and homogenized at 1173 K for several hours, and then the crystal growth was fostered by slowly cooling down the temperature from 983 K to 913 K in 560 h in a temperature gradient of roughly 1 K cm$^{-1}$. Then the grown rod was fast cooled to room temperature. In order to obtain crystals with high-quality, we performed surface cleaning for all starting materials to remove the oxide layers formed in air.[37, 38] Scanning transmission electron microscopy (STEM) measurements are performed on a high-quality (PbSe)$_5$(Bi$_2$Se$_3$)$_6$ single crystal. After fully grounding the single crystal in acetone, the small fragments suspended in acetone were dripped onto a TEM microgrid. The atomic positions of the (PbSe)$_5$(Bi$_2$Se$_3$)$_6$ were characterized using an ARM-200CF (JEOL, Tokyo, Japan) transmission electron microscope operated at 200 kV.

High-pressure electric resistance measurements of (PbSe)$_5$(Bi$_2$Se$_3$)$_6$ crystals were performed by the van der Pauw four-probe method with NaCl as pressure transmitting medium as described elsewhere[21, 39, 25]. Cubic BN and epoxy mixture were used as an insulation layer. Platinum leads were arranged in nonmagnetic copper beryllium (BeCu) diamond anvil cell. The pressure was calibrated by the ruby measurements at room temperature.[40] A magnet-cryostat (Dynacool, Quantum Design, $T_{min}$ = 1.8 K) was used to take the cryogenic setup and magnetic field measurements.

High-pressure *in-situ* Raman spectroscopy investigation on (PbSe)$_5$(Bi$_2$Se$_3$)$_6$ was carried out on a Raman spectrometer (Renishaw in Via, U.K.) with a laser excitation wavelength of 532 nm as well as a low-wavenumber filter. Symmetric DAC with anvil culet sizes of 300 μm and mineral oil was used as pressure transmitting medium. The pressure was calibrated by the ruby measurements at room temperature.[40]

The pressure dependence of the XRD pattern was measured at beamline 15U at Shanghai Synchrotron Radiation Facility (x-ray wavelength λ = 0.6199 Å). Symmetric diamond anvil cell (DAC) with anvil culet sizes of 300 μm, and T301 gaskets were used. Neon and Mineral oil were used as the pressure transmitting medium (PTM), and pressure was determined by the ruby luminescence method in the lower-pressure experiment.[40] Rietveld refinements of the crystal structures under high pressure were performed using the General Structure Analysis System (GSAS) and graphical user

interface EXPGUI package.[41]

ACKNOWLEDGMENT

We thank Prof. Yanming Ma for valuable discussions. This work was supported by the National Natural Science Foundation of China (Grant No. 12004252, U1932217, 11974246, 52072400, 52025025, 92065109), the National Key R&D Program of China (Grant No. 2018YFA0704300, 2021YFA1401800, 2018YFE0202601, 2020YFA0308800 and 2022YFA1403400), Shanghai Science and Technology Plan (Grant No. 21DZ2260400), and Beijing Natural Science Foundation (Z190010, Z210006, Z190006). The authors thank the support from Analytical Instrumentation Center (# SPST-AIC10112914), SPST, ShanghaiTech University. The authors thank the staffs from BL15U1 at Shanghai Synchrotron Radiation Facility for assistance during data collection.